\documentclass[a4paper,11pt]{article}
\usepackage{jheppub}
\usepackage{textcase}
\usepackage{graphicx,epsfig,psfrag,amssymb,hyperref}
\usepackage{multirow}
\usepackage{color,graphicx,epsfig,psfrag,amsmath,empheq}
\usepackage{bm}
\usepackage{mathrsfs,amsfonts,soul,color}
\usepackage[caption=false]{subfig}
\usepackage{slashed}
\usepackage{placeins}
\usepackage[utf8]{inputenc}

\newcommand{\cL}{\mathcal{L}}

\newcommand{\cO}{\mathcal{O}}

\newcommand{\eg}{\textit{e.g.}}

\newcommand{\MeV}{{\rm MeV}}
\newcommand{\TeV}{{\rm TeV}}
\newcommand{\GeV}{{\rm GeV}}

\newcommand{\be}{\begin{equation}}
\newcommand{\ee}{\end{equation}}
\newcommand{\bea}{\begin{eqnarray}}
\newcommand{\eea}{\end{eqnarray}}

\title{
A Custodial Symmetry for Muon $g-2$}
\preprint{LAPTH-016/21}

\author[a]{Reuven Balkin,} 
\author[b]{C\'edric Delaunay,}
\author[c]{Michael Geller,} 
\author[a]{Enrique Kajomovitz,} 
\author[d]{Gilad Perez,}
\author[d]{Yogev Shpilman,}
\author[a]{and Yotam Soreq}

\affiliation[a]{Physics Department, Technion -- Israel Institute of Technology, Haifa 3200003, Israel}
\affiliation[b]{LAPTh, CNRS -- USMB, BP 110 Annecy-le-Vieux, F-74941 Annecy, France}
\affiliation[c]{Department of Physics, Tel Aviv University, Tel Aviv, Israel}
\affiliation[d]{Department of Particle Physics and Astrophysics, Weizmann Institute of Science, Rehovot 7610001, Israel}

\emailAdd{reuven.b@campus.technion.ac.il}
\emailAdd{cedric.delaunay@lapth.cnrs.fr}
\emailAdd{micgeller@tauex.tau.ac.il}
\emailAdd{enrique@physics.technion.ac.il}
\emailAdd{gilad.perez@weizmann.ac.il}
\emailAdd{yogev.shpilman@weizmann.ac.il}
\emailAdd{soreqy@physics.technion.ac.il}

\abstract{
We discuss the recent results on the muon anomalous magnetic moment in the context of new physics models with light scalars. 
We propose a model in which the one-loop contributions to $g-2$ of a scalar and a pseudoscalar naturally cancel in the massless limit due to the symmetry structure of the model. 
This model allows to interpolate between two possible interpretations. 
In the first interpretation, the results provide a strong evidence of the existence of new physics, dominated by the positive contribution of a CP-even scalar. 
In the second one, supported by the recent lattice result, the data provides a strong upper bound on new physics, specifically in the case of (negative) pseudoscalar contributions. 
We emphasize that tree-level signatures of the new degrees of freedom of the model are enhanced relative to conventional explanations of the discrepancy. As a result, this model can be tested in the near future with accelerator-based experiments and possibly also at the precision frontier.
}

\begin{document}
\maketitle

\section{Introduction}
\label{sec:Intro}

Recently, the Muon g-2 experiment at Fermilab has announced its first preliminary result~\cite{Abi:2021gix,Albahri:2021kmg,Albahri:2021ixb}
\begin{align}
    a_\mu^{\rm FNAL} 
=   116592040(54)\times10^{-11}\,,
\end{align}
where $a_\mu \equiv (g_\mu -2 )/2$ is the magnetic moment anomaly of the muon.  
Averaging it with the BNL result~\cite{Bennett:2006fi} leads to a new world average~\cite{Abi:2021gix} 
\begin{align}
    a^{\rm WA}_\mu 
=   116592061(41)\times 10^{-11}\,.
\end{align}
When compared to the standard model (SM) theoretical prediction~\cite{Aoyama:2020ynm} $a_\mu^{{\rm SM},{R}}=116591810(43)\times10^{-11}$, one finds~\cite{Abi:2021gix}
\begin{align}
    \label{eq:aR}
    \Delta a_\mu^{R} 
    \equiv
    a^{\rm WA}_\mu - a^{{\rm SM},{R}}_\mu
=   251(59)\times10^{-11}\,,
\end{align}
which is $4.2$ standard deviations~($\sigma$) from zero. 
However, we can also compare $a^{\rm WA}_\mu$ to the SM prediction based the leading order hadronic vacuum polarization~(HVP) estimated by lattice QCD calculations. 
Reference~\cite{Borsanyi:2020mff} reports $a^{\rm HVP-LO,lattice}_\mu=707(55)\times10^{-11}$, which translate to  $a^{\rm SM,lattice}=11659195163(58)\times10^{-11}$.
Then, the resulting difference between the SM and $a^{\rm WA}_\mu$ is
\begin{align}
    \label{eq:alattice}
    \Delta a^{\rm lattice}_\mu 
    \equiv
    a^{\rm WA}_\mu - a^{\rm SM,lattice}_\mu
=   109(71)\times10^{-11}\,,
\end{align}
which is only $1.6\,\sigma$ from zero. 

One can interpret this result in two ways. The first interpretation is that we have a substantial evidence of new physics~(NP). Most of the literature focusing on this direction is dedicated to investigating what type of SM extension could account for the difference between the experimental result and the theoretical prediction, see for example~\cite{Miller:2012opa,Battaglieri:2017aum,Strategy:2019vxc,Zyla:2020zbs}. In this interpretation, the $2\,\sigma$ preferred range deduced from Eq.~\eqref{eq:aR} is $133 < a^{\rm NP}_\mu/10^{-11} < 369$.
Alternatively, one can interpret the data in view of the new lattice result of Ref.~\cite{Borsanyi:2020mff}, which provides a bound on new states coupled to muons. In this case, the allowed $2\,\sigma$ range from Eq.~\eqref{eq:alattice} yields $-33 < a^{\rm NP}_\mu/10^{-11} < 251$. The model proposed in this letter allows us to interpolate between the above two different interpretations of the result.

We propose a model where the one-loop contribution to the anomalous magnetic moment of the muon is naturally suppressed in the limit where the new scalar states are massless~\cite{toappear}. This cancellation allows the model to evade the bound of Eq.~\eqref{eq:alattice}, especially for NP masses much lighter than the muon mass. Alternatively, a small deformation of the model could generate a large enough contribution to $a_\mu$, thus accomodating the strong evidence of NP in Eq.~\eqref{eq:aR}.

Despite this one-loop suppression of $a_\mu$, other tree-level processes, such as bremsstrahlung emission of light NP states or correction to bound-state energy levels, are unsuppressed. As a result, the relevant rates are enhanced by a factor as large as $\cO\left(100\right)$ and become visible. Searches which probe the NP coupling to muons are therefore reviewed in this context. We also discuss implications of possible coupling of the new scalar states to electrons, assuming a naive $m_e/m_\mu$ suppression factor relative to muons.

\section{The model}
\label{sec:model}

The basic observation behind our construction stems from the fact that the one-loop contribution to $(g-2)_\mu$ of a massless scalar, $\phi$ and pseudoscalar, $\pi$ with sole Yukawa coupling to muons, respectively $y_\phi$  and $y_\pi$, are 
\begin{align}
    a_\mu^{\phi}  
=   \frac{3y_\phi^2}{16\pi^2} \,, 
    \qquad
    a_\mu^\pi=- \frac{y_\pi^2}{16\pi^2} \,.
\end{align}
We thus learn that in the massless limit a scalar can account for the central value of the anomaly provided that 
\begin{align}
    y_\phi\approx 3.6 \times 10^{-4}\,.
\end{align}
A single pseudoscalar cannot account for the anomaly as its contribution is of opposite sign. 
Therefore, the 95\,\% CL bound on its coupling is 
\begin{align}
    y_\pi\lesssim 2.3 \times 10^{-4}\,.
\end{align}

We further observe that in a model consisting of three massless pseudoscalars $\pi_i$ ($i=1..3$) and a single massless scalar with equal Yukawa couplings, $y_\phi=y_{\pi_i}$, the one-loop contribution to the anomalous magnetic moment of the muon vanishes~\cite{toappear}.\\

We now introduce a natural model which realizes the above cancellation. 
Consider the two fourplet fields,
\begin{align}
    \varepsilon
=   \begin{pmatrix}
        \ell_{1}\\
        \ell_{2}\\
        \ell_{3}\\
        \ell_{4}
    \end{pmatrix}\,,  
    \qquad\quad
    \sigma
=   \begin{pmatrix}
        \phi\\
        \pi_{1}\\
        \pi_{2}\\
        \pi_{3}
    \end{pmatrix} \, ,
\end{align}
as an extension to the SM, where $\ell_{i}$ are Weyl spinors, and $\sigma$ consists of four real scalars. For reasons that will become clear below we define the components of the $\sigma$ field as one scalar, $\phi$ and three pseudoscalars, $\pi_{i}$\,. 
The Yukawa and mass terms of the fields in the Lagrangian are (using two-component spinor notations for convenience)~\footnote{For the sake of simplicity we only show the relevant terms in the broken electroweak-symmetry phase. We further assume alignment in the leptonic flavor space and only coupling to muons.}
\begin{align}
  -  \cL
&   \supset
    \frac{1}{2}m^2_\sigma \, \sigma^{T}\cdot \sigma
    +\left( \sigma^{T}\cdot\hat{y}\cdot\varepsilon\, \,\ell^{c}
    +m^\dagger\cdot\varepsilon\,\,\ell^{c}
    +\text{h.c.} \right)\,,
\end{align}
where the matrix $\hat{y}\propto \langle H\rangle/\Lambda$ and the vector $m\propto y_{\rm SM} \langle H\rangle$ are treated here as spurions,
$H$ being the SM-Higgs field, $\Lambda$ the effective cutoff of the model and $y_{\rm SM}$ the SM lepton Yukawa coupling. The field $\ell^{c}$ is the SM left-handed~(LH) lepton doublet, while 
the right-handed~(RH) SM singlet is a linear combination of the $\ell_i$ fields given by $m^\dagger\cdot\varepsilon$. We implicitly assume that the other three orthogonal combinations which we denote $\psi_{1,2,3}$ acquire large Dirac masses by coupling to additional singlet fields $\psi^c_{1,2,3}$ and are therefore decoupled. 
In addition, we assume CP symmetry. 

The Lagrangian has a global symmetry $G\equiv\rm{O}\left(4\right)_{\sigma}\times {\rm U}\left(4\right)_{\varepsilon}\times {\rm U}(1)_{\ell^c}$ where the fields transform as
\begin{align}
    \sigma&\rightarrow O\cdot \sigma\,,
    &\varepsilon&\rightarrow U \cdot\varepsilon\,,
    &\ell^c&\rightarrow e^{i\alpha} \ell^c \, ,
\end{align}
and the spurions as
\begin{align}
    \hat{y}&\rightarrow e^{-i\alpha}\,O\cdot  \hat{y} \cdot  U^{\dagger}  \,,
    & m\rightarrow e^{i\alpha}U\cdot m \,,
\end{align}
with $O$, $U$ and $e^{i\alpha}$ denoting ${\rm O}\left(4\right)_{\sigma}$, ${\rm U}\left(4\right)_{\varepsilon}$ and ${\rm U}(1)_{\ell^c}$ transformations, respectively. 
Generically $\hat y$ breaks $G$ to a common ${\rm U}(1)_\mu$ associated with the conserved muon number. Here, instead, we assume that the background value of $\hat{y}$ is not generic but leads to the pattern of symmetry breaking
\begin{align}
    \label{eq:yhatSymBreak}
    G \, \rightarrow \, 
    {\rm O}\left(4\right)_{\sigma+\varepsilon}\times {\rm U}\left(1\right)_\mu\,.
\end{align}
  Without loss of generality, the background value of $m$ leads to
\begin{align}
    G \, \rightarrow \, 
    {\rm O}\left(4\right)_{\sigma}\times {\rm U}\left(3\right)_{\varepsilon}\times {\rm U}\left(1\right)_\mu\,,
\end{align}
where the ${\rm U}(3)_\varepsilon$ symmetry corresponds to the residual freedom of transforming the three linear combinations of $\ell_i$ orthogonal to $m^\dagger \cdot \varepsilon$. Given the above pattern of symmetry breaking the spurions can be brought to  $\langle \hat y\rangle = y_\ell\times {\rm diag}(1,1,1,1)$ and $\langle m\rangle = m_l\times (e^{i \theta_1},e^{i \theta_2},e^{i\theta_3},e^{i\theta_4})^T$ where $y_\ell$, $m_\ell$ are real and the phases $\theta_{1,2,3,4}$ are either zero or $\pi/2$ under the assumption that CP is conserved.

There are then two possibilities: either all the phases are the same or one of them (at least) is $\pi/2$. In the first case, the unbroken symmetry in the presence of both spurions is ${\rm O}\left(3\right)_{\sigma+\varepsilon}\times {\rm U}\left(1\right)_\mu$, since it is always possible to have $\langle m\rangle\propto (1,0,0,0)^T$ using an O(4) transformation. In this limit only one component of $\sigma$ interacts with muons (be it a scalar or pseudoscalar depends on the relative phase between the spurions) with no cancelling counterpart. In the second case the residual symmetry is only $O\left(2\right)_{\sigma+\varepsilon}\times U\left(1\right)_\mu$ since the relative phase is not removable by O(4) transformations which can only yield $\langle m\rangle\propto (1,i,0,0)^T$. In this limit there are two states coupled to muons, one scalar and one pseudoscalar, whose interactions (and masses) are fixed by symmetries to enforce cancellation between their one-loop contributions to $(g-2)_\mu$. 

Henceforth we focus on the O(2)-symmetric case and work in a basis where the background values of the spurions are\label{eqn:O2structure}
\begin{align}
    \left\langle \hat{y}\right\rangle 
=   \frac{y_\ell}{2}
    \begin{pmatrix}
        1\\
        & i\\
        &  & i\\
        &  &  & i
    \end{pmatrix} \,, 
    \qquad\quad
    \left\langle m\right\rangle 
=   \frac{m_\ell}{2}\begin{pmatrix}
        1\\
        1\\
        1\\
        1
    \end{pmatrix}\,,
\end{align}
where $y_\ell$ and the lepton mass $m_\ell$ are real and positive. In the $m\to0$ limit we find that the combination $\ell'\equiv \frac{1}{2}\sum_{i=1}^{4} \ell_i$, remains massless. 
This state will couple to its SM, LH doublet, partner field $\ell^c$ through the Higgs field, and would correspond to the SM RH muon field. 

Note that the above structure is technically natural and is protected by collective breaking of the above symmetries. If only one of the spurion is present, the muon field couples at most to one scalar state. We also assume that $\sigma$ has vanishing vacuum expectation value, so it does not affect the mass of the SM lepton. 

The relevant part of the low-energy Lagrangian (below the $\mathcal{O}($TeV) mass of the $\psi_i$'s) is
\begin{align}
    \label{eq:Yukawa_sector}
   - \cL_{\rm eff}
&=  y_\ell \,\phi\left(\ell'\ell^{c}+\left(\ell'\ell^{c}\right)^{\dagger}\right)
    +\sqrt{3}i\,y_\ell \,\pi\left(\ell'\ell^{c}-\left(\ell'\ell^{c}\right)^{\dagger}\right)+m_\ell \ell'\ell^{c}
\end{align}
where
\begin{align}
    \pi
\equiv   \frac{1}{\sqrt{3}}\left(\pi_{1}+\pi_{2}+\pi_{3}\right)\,.
\end{align}
The field $\pi$ is the ``custodian'' field whose one-loop contribution to $g-2$ is forced by the O(2) symmetry to cancel that of $\phi$. The two linear combinations orthogonal to the custodian remain non-interacting. Note that these states are stable due to the O(2) symmetry and may be valid dark matter candidates. A detailed analysis is beyond the scope of the current work.
We expect a small deviation from unity in the scalar-to-pseudoscalar coupling ratio $y_{\phi}/y_{\pi}$  due to different RG flow from the TeV scale, where they are equal, to the MeV scale of the measurements. However, such effects are $\cO\left(y_{\ell}^4\right)$ and can be safely neglected. 

Finally, one can introduce a small soft O(4) breaking (yet O(2) preserving) term which lifts the degeneracy between $\phi$ and $\pi$ masses (respectively, $m_\phi$ and $m_\pi$). In order to interpolate between the different limits we define
\begin{align}
    \Delta m \equiv m_\pi- m_\phi \, ,  \qquad\quad 
    \Sigma m \equiv m_\pi + m_\phi\,,
\end{align}
and we denote the degenerate scalar and pseudoscalar masses by $m_\sigma$ when $\Delta m=0$.\\

The above construct can arise from extra-dimensional (alternatively 4-dimensional four-site) models~\cite{toappear} consisting of a fourplet of bulk fermions with only one zero mode identified with $\ell'$, the massive orthogonal combinations being KK excitations. The structure of the spurions is Eq.~\eqref{eqn:O2structure} can be realized by imposing appropriate boundary conditions for the fields propagating in the bulk.  
Finally, we note that in order to keep the scalar masses light, say $m_\sigma\lesssim\,10\,\MeV$ requires new physics to appear at a scale $4\pi\, m_\sigma/y_\ell\sim\, \TeV$, which is consistent with the scale of the new states in our construction.

\section{Phenomenological implications}
\label{sec:treelevel}

\begin{figure}[t]
    \includegraphics[width=0.95\textwidth]{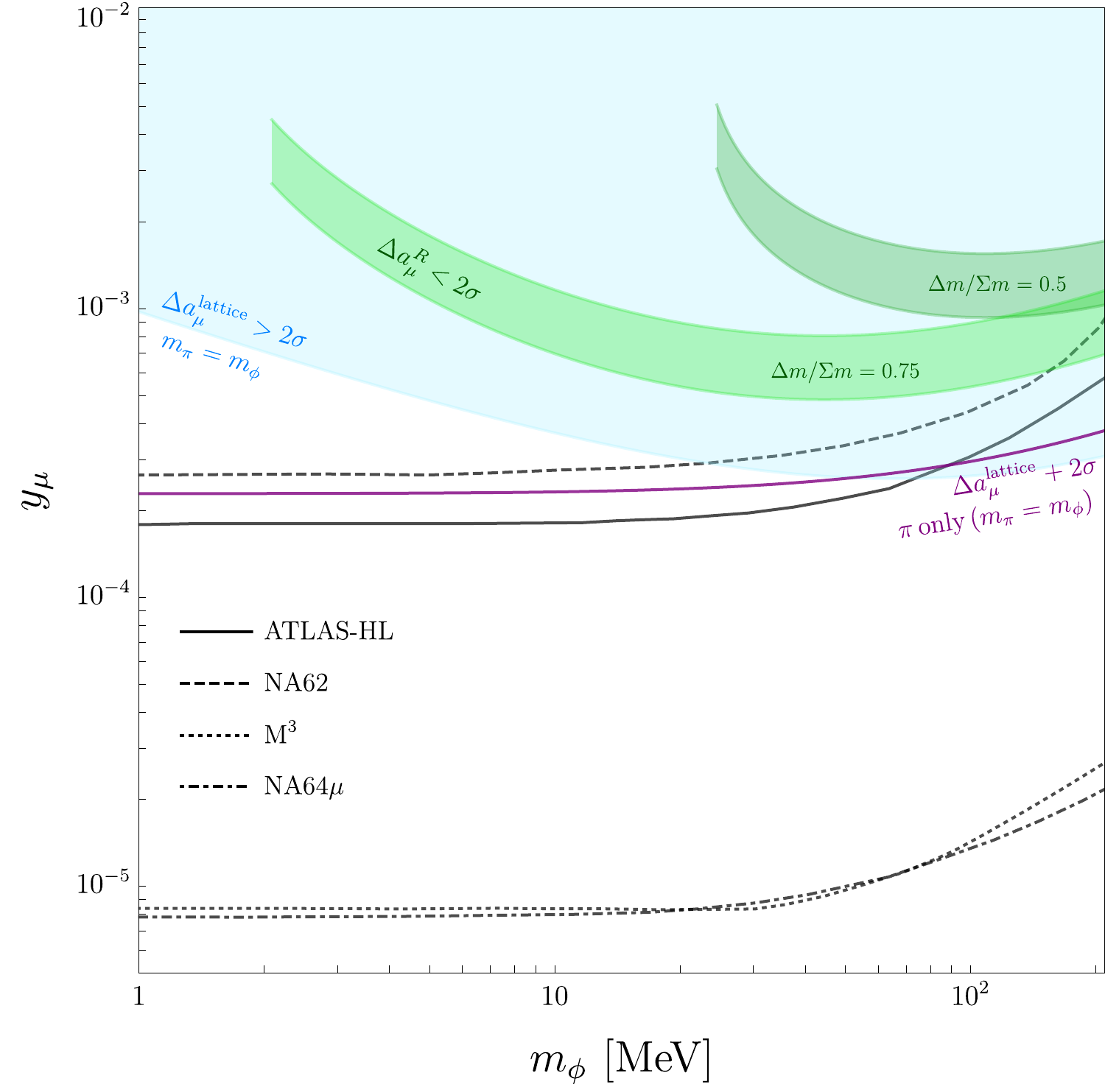}
    \caption{Parameter space of muonphilic scalars compared to experimental sensitivities of tree-level based searches. The green bands are preferred regions interpreting the $(g-2)_{\mu}$ result as evidence of new physics, $| \Delta a_{\mu}^{\rm{NP}} - \Delta a_{\mu}^{\rm{R}} |< 2 \sigma^{\rm{R}}$. They are cut off where the estimated two-loop contribution reaches $10\%$ of the one-loop contribution, see Eq.~\ref{eq:amuNP}.
    The blue area is the excluded region interpreting the $(g-2)_{\mu}$ result as a bound on new physics, $ \Delta a_{\mu}^{\rm{NP}} -  \Delta a_{\mu}^{\rm{lattice}} > 2\sigma^{\rm{lattice}}$. The purple line indicates the $a_{\mu}^{\rm NP} < 0$ bound on a single pseudoscalar state, $\Delta a_{\mu}^{\rm{lattice}} - \Delta a_{\mu}^{\rm{NP}} >  2\sigma^{\rm{lattice}}$. The dashed lines represent the sensitivity of the ATLAS-HL, NA62, M$^3$ and NA64$\mu$ experiments.}
    \label{fig:mucoupling}
\end{figure}

For $m_\sigma \ll m_\mu$, the one-loop contribution to the muon magnetic moment is suppressed and  vanishes in the limit of $m_\sigma\to0$.  
However, this exact cancellation is lifted by two effects. 
The first one arises from the finite mass of the $\sigma$ field. It is straightforwardly evaluated from the one-loop functions of scalar and pseudoscalar, see \eg~\cite{Queiroz:2014zfa}. 
The second one is due to two-loop contributions whose ratio differs from that of the one-loop contributions in the massless limit. We estimate the two-loop correction to be of the order of a single pseudoscalar one-loop contribution times an additional suppression of $\alpha/4\pi\sim 10^{-3}$. 
In total, the contribution to $a_\mu^{\rm NP}$ at the small mass limit, $m_\sigma \ll m_\mu$, can be estimated as 
\begin{align}
    \label{eq:amuNP}
    a_\mu^{\rm NP} 
    \approx
    \frac{y^2_\mu}{8\pi^2}\left( C_{2}\frac{\alpha}{4\pi} -  \pi\frac{m_\sigma}{m_\mu}\right) \, ,
\end{align}
where $C_2\sim\mathcal{O}(1)$ with no fixed sign. In practice, the two-loop contribution is sub-leading in most of the parameter space of interest. 
In particular, the above estimate is at most an $\cO(1\%)$ correction to the one-loop result for $m_\sigma=1\,\MeV$ in the O(2) symmetric limit. 
For $\Delta m/\Sigma m\neq0$, the two-loop contribution becomes important only for peculiar values of $m_\phi$ where the scalar and pseudoscalar contributions cancel out, see Fig.~\ref{fig:mucoupling}.   

As mentioned above, the degeneracy between the scalar and pseudoscalar is lifted for $\Delta m \ne0$.
For $\Delta m > 0$, $a^{\rm NP}_\mu$ is dominated by $\phi$ and positive, and can account for the central value in Eq.~\eqref{eq:aR}. 
Conversely, for $\Delta m \le 0$, $a^{\rm NP}_\mu$ is always negative, preventing an explanation of the the central value of Eq.~\eqref{eq:aR}.  
We demonstrate this in Fig.~\ref{fig:mucoupling}, with the $2\,\sigma$ excluded region (in blue) from Eq.~\eqref{eq:alattice}, in the O(2) symmetric limit.
The green bands account for the central value of the discrepancy in Eq.~\eqref{eq:aR}, for the special cases of $\Delta m/\Sigma m = 0.5$ and $0.75$. 

From Eq.~\eqref{eq:amuNP}, we learn that the strong bound from $(g-2)_\mu$ is weakened in our model thanks to the (partial) cancellation of the scalar and pseudoscalar contributions. 
This is illustrated in Fig.~\ref{fig:mucoupling} with the $\Delta a^{\rm lattice}_\mu$ bound on a single pseudoscalar state (purple). This region of parameter space between the purple line and the blue area is currently unexplored by $(g-2)_\mu$ but is possibly probed by tree-level searches for muon force carrier~(MFC) as we show below. 
Among these, we find muon beam experiments~\cite{Chen:2017awl,Kahn:2018cqs,Chen:2018vkr},  ATLAS~\cite{Galon:2019owl}, FASER$\nu$~\cite{Fnumu}, $K\to\mu\nu$ decays at NA62~\cite{Krnjaic:2019rsv}, and $B$-factories~\cite{Jho:2019cxq} have promissing prospects. 
As the cancellation is only effective for pseudoscalar/scalar masses below $m_\mu$, it will not affect the relevant parameter space for MFC searches which focus on dimuon scalar decay as in Refs.~\cite{TheBABAR:2016rlg,Batell:2016ove}. 
In addition, we consider precision measurements such as muonium~(Mu) positronium~(Ps) and helium spectroscopy, and that of the electron anomalous magnetic moment.

\subsection{Muonphilic tree-level based searches}
\label{sec:mucoupling}

We start by exploring the simple and minimal muonphilic case, in which the $\sigma$ field only couples to  muons and constitutes a new MFC. 
Most of the analyses provided in the various proposals for MFC searches assume a single field (usually a scalar or a vector) as the new MFC. 
Our goal in this section is to estimate the effect of the additional pseudoscalar on the different experimental signatures and the associated projected sensitivities. 
 
First, we consider searches where a muon propagates through a fixed target emitting the MFC in a bremsstrahlung-like process. 
We focus on signatures in which the MFC itself does not leave any trace in the detector either because it is long lived (which is the natural expectation for masses below the dimuon threshold) or decays predominantly to particles in a secluded sector (\eg~dark matter particles). 
There are two main possible observables related solely to the muon momentum that arise due to the scalar-emission: 
(i)~missing muon momentum, see the M$^3$ experiment in Fermilab~\cite{Kahn:2018cqs}, ATLAS~\cite{Galon:2019owl} and NA64$\mu$~\cite{Gninenko:2014pea,Banerjee:2016tad,Gninenko:2016kpg,Chen:2018vkr} experiments (see also~\cite{Izaguirre:2014bca,Mans:2017vej} for the electron case);
(ii)~change in the muon's direction, a kink in its trajectory. 

The second class of observable, associated with the kink, are detectable with high precision in emulsion targets, see \eg~\cite{Kodama:2000mp,Agafonova:2019wvk}. 
The FASER/FASER$\nu$ experiment~\cite{Feng:2017uoz,Ariga:2018pin,Abreu:2019yak} has an excellent spacial resolution and allows the detection of kinks in muon tracks at $\cO(10^{-4}\,{\rm rad})$ level~\cite{Abreu:2020ddv}. 
This is a powerful probe of MFC that can cover a large part of MFC parameter space of $y_\mu\sim\cO(10^{-3}-10^{-4})\,$   
during the first run of FASER$\nu$ (LHC Run-3)~\cite{Fnumu}. 

We have estimated the corresponding production processes in the $m_\sigma \ll m_\mu$ limit, using the 
Weizs\"cker-Williams approximation~\cite{vonWeizsacker:1934nji,Williams:1935dka}. 
We find that the total cross section for a scalar is  similar to that of the pseudoscalar in the limit of equal couplings.
However, using the analysis of Ref.~\cite{Liu:2016mqv,Liu:2017htz} we do observe that the pseudoscalar differential rate tends to be harder than that of the scalar, which we ignore in our recast. 
Thus, below we present the projected bounds using the scalar-only rates of~\cite{Galon:2019owl,Kahn:2018cqs,Chen:2018vkr} as a conservative estimation for the sensitivity in our model. 

A different approach is taken in~\cite{Krnjaic:2019rsv}, where the MFC is probed through the $K \to \mu\nu X$ decays in NA62. 
When comparing the exclusive branching ratios we found (in the $m_\sigma\to0$ limit)
\begin{align}
   \frac{\Gamma(K\to \mu\nu \, \phi)}{\Gamma(K\to \mu\nu\, \Pi)}
   \sim 2\,,
\end{align}
where a missing energy cut of $E^2_{\text{\tiny miss}}>0.05\,\GeV^2$ was imposed. (See Ref.~\cite{Krnjaic:2019rsv} for further details.) 
We reach a similar conclusion regarding this search, namely for low masses the signal is dominated by the scalar and the current projected sensitivities for a single scalar constitute a conservative estimation for the projected sensitivities of our model.

These projections are presented as black lines in Fig.~\ref{fig:mucoupling}, showing that M$^3$, NA62, NA64$\mu$ and ATLAS experiments can probe unexplored parameter space, in the context of either of the two possible interpretations of the  $(g-2)_\mu$ results. 
Note that in all these searches, we expect that a full analysis (to be performed elsewhere) would result in an $\cO(1)$ deformation of the current projected sensitivities for a single scalar.

Finally, the $\mu^+\mu^-$ pair can form a bound state, denoted as
true-muonium~(TM), which has not been observed experimentally yet.
However, it would serve as an important alternative probe of our and other MFC models~\cite{CidVidal:2019qub} once its properties are measured precisely. 

\subsection{Coupling to electron and muon}
\label{sec:emucoupling}

A possible coupling of $\sigma$ to electrons would provide additional ways to probe our model.
We assume the following relation among the coupling, $y_e=y_\mu \times (m_e/m_\mu)$, for definiteness. 

\begin{figure}[t]
    \includegraphics[width=0.95\textwidth]{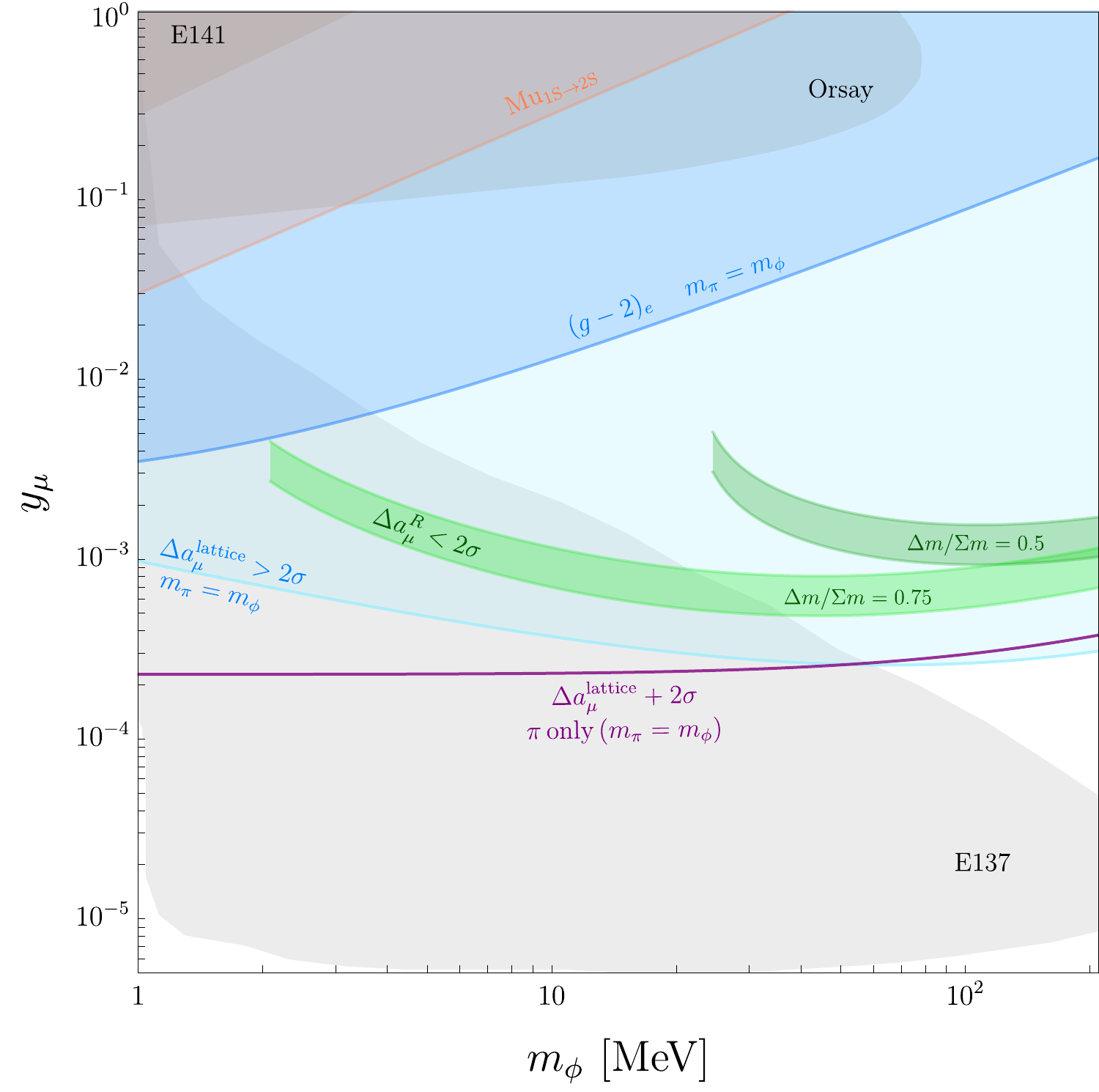}
    \caption{Parameter space of scalars with electron and muon couplings. The green bands are the preferred regions interpreting the $(g-2)_{\mu}$ result as evidence of new physics, $| \Delta a_{\mu}^{\rm{NP}} - \Delta a_{\mu}^{\rm{R}} |< 2 \sigma^{\rm{R}}$. They are cut off where the estimated two-loop contribution reaches $10\%$ of the one-loop contribution, see Eq.~\ref{eq:amuNP}. 
    The light blue area is the excluded region interpreting the $(g-2)_{\mu}$ result as a bound on new physics, $ \Delta a_{\mu}^{\rm{NP}} -  \Delta a_{\mu}^{\rm{lattice}} > 2\sigma^{\rm{lattice}}$. The dark blue area shows the excluded region from the $(g-2)_{e}$ measurement. The purple line indicates the $a_{\mu}^{\rm NP} < 0$ bound on a single pseudoscalar state, $\Delta a_{\mu}^{\rm{lattice}} - \Delta a_{\mu}^{\rm{NP}} >  2\sigma^{\rm{lattice}}$. The gray regions are excluded by tree-level based searches at the Orsay, E137 and E141 electron beam dump experiments. The orange region is excluded by the measurement of the 1S-2S transition frequency of muonium.}
    \label{fig:muecoupling}
\end{figure}

The phenomenology of this model, associated with tree-level based searches, is similar to the model of leptonic Higgs protal~\cite{Batell:2016ove}.
At the intensity frontier, we consider bounds from the electron beam dumps E141~\cite{Riordan:1987aw}, Orsay~\cite{Davier:1989wz} and E137~\cite{Bjorken:1988as} and adopt the recast of Refs.~\cite{Batell:2016ove,Liu:2016mqv,Liu:2017htz}. 
As illustrated in Fig.~\ref{fig:muecoupling}, only Orsay and E137 are relevant to our model. 
We note that these bounds can be avoided in non-minimal models where the dominant $\phi$ and $\pi$ decay mode is invisible. 

At the precision frontier, the muonium (the $\mu^+e^-$ bound state) is a dedicated candidate for testing the existence of a light force carrier between muon and electron, see~\cite{Frugiuele:2019drl}. 
The most accurate frequency measurement is for the 1S-2S transition with $\nu_{\rm 1S-2S}^{\rm Mu}({\rm exp})=2455528941.0(9.8)\,$MHz~\cite{Meyer:1999cx}.
This is in a good agreement with the QED theory prediction $\nu_{\rm 1S-2S}^{\rm Mu}({\rm th})= 2455528935.4(1.4)\,$MHz~\cite{Pachucki:1996jw,Karshenboim:1997zu,Meyer:1999cx}, whose principal uncertainty is due to the muon-to-electron mass ratio~\cite{Mohr:2015ccw}, currently extracted from measurements of Zeeman transitions within the muonium ground state~\cite{Liu:1999iz}. 
The 1S-2S transition will be soon remeasured by the MuMASS collaboration at PSI with $\cO({\rm kHz})$ precision~\cite{Crivelli:2018vfe}. 
In the meantime the theory prediction is expected to improve by about one order of magnitude thanks to a new measurement of the 1S hyperfine splitting at J-PARC~\cite{Strasser:2019fbk}. 
The Mu 1S-2S transition is mostly sensitive to spin independent force from the scalar exchange (the pseudoscalar induce spin-dependent force, which is highly suppressed). The resulting upper bound of $y_\mu$ from the difference between the theory calculation and the exprimental result is plot as the orange curve in Fig.\ref{fig:muecoupling}. As we can see it less sensetive than other probes. 
We note that for $m_\sigma < 0.1\,\MeV$ (not shown in the plot) the bounds from $(g-2)_{e,\mu}$ and from electron beam dump become weaker than the one obtained from Mu spectroscopy.

The positronium (the $e^+e^-$ bound state) and helium atom as well as the electron magnetic moment, $a_e$, are sensitive to electrophilic forces.
The 1S-2S transition frequency in ortho-Ps (the spin triplet configuration) has been measured with MHz accuracy~\cite{Fee:1993zza} and agrees well with theory~\cite{Pachucki:1997vm} which is limited by unknown three-loop corrections in QED. 
Several helium 4 transitions are measured with kHz accuracy with good agreement with theory calculation at the MHz level~\cite{Pachucki:2017xcg}. 
The sensitivity to electron-electron new physics interactions is comparable to positronium~\cite{Delaunay:2017dku}.  
We find that the resulting bounds from precision positronium and helium spectroscopy are weaker than other bound muonium. This is because we assume that the electron coupling is suppressed by $m_e/m_\mu$.

Free electron $(g-2)_e$ is measured at $\cO(10^{-11})$ accuracy in Penning trap experiments $a_e({\rm exp})=0.001\, 159\, 652\, 180\, 73 (28)$~\cite{Hanneke:2008tm}, which requires a five-loop calculation in QED~\cite{Aoyama:2012wj} for a meaningful comparison to theory. 
The dominant theory uncertainty comes from $\alpha$ which is currently best extracted from a combination of fundamental constants including the Rydberg constant, the electron and Rubidium atomic masses and the ratio of the Planck constant to the Rubidium mass recently measured at LKB~\cite{Morel:2020dww}, yielding $a_e({\rm th})=0.001\,159\, 652\,180\,252(95)$. 
The resulting bound from $(g-2)_e$ is plotted in Fig.~\ref{fig:muecoupling}.

\section{Conclusions}
\label{sec:conclusions}

We investigate the new physics implications of the recent measurements of the muon magnetic moment in a model with new light scalar states. The symmetry of the model ensures that the one-loop contribution of these states to the  muon anomalous magnetic moment naturally vanishes in the massless limit.
In the muonphilic case, we find that near future searches based on tree-level scalar emission at the luminosity frontier are able to cover most of the relevant parameter space of the model. 
Assuming that the new states couple to the electron as well, we find that the model is further probed by electron beam-dump experiments and muonium spectroscopy at the precision frontier. 

\section*{Acknowledgements}
We thank Yossi Nir for a comment made that drove us to think about a mechanism to suppress the one-loop contributions to $g-2$, without affecting the tree level ones.
We thank Iftah Galon and Omer Tsur for useful discussions. 
MG is supported by grants from BSF, ISF and GIF foundation.
EK is supported by grants from BSF and ISF. 
The work of GP is supported by grants from BSF-NSF, the Friedrich Wilhelm Bessel research and award, GIF, ISF, Minerva and Weizmann's SABRA - Yeda-Sela - WRC Program, the Estate of Emile Mimran, and The Maurice and Vivienne Wohl Endowment.
YS and RB are supported by grants from NSF-BSF, ISF and the Azrieli foundation. 
YS is a Taub fellow (supported by the Taub Family Foundation).


\bibliographystyle{JHEP}
\bibliography{ref}

\end{document}